\documentclass[pla,twocolumn,showpacs,showkeywords,preprintnumbers,amsmath,amssymb]{revtex4}
\usepackage{tipa}
\usepackage{amssymb}
\usepackage{txfonts}

\usepackage{hyperref}
\usepackage{graphicx}
\usepackage{epsfig}
\begin{document}

\title{Remote generation of entanglement for individual atoms via optical fibers}
\author{Y. Q. Guo\footnote{
Corresponding author: yqguo@newmail.dlmu.edu.cn}, H. Y. Zhong, Y.
H. Zhang} \affiliation{Department of Physics, Dalian Maritime
University, Dalian 116026, P.R.China}

\author{H. S. Song} \affiliation{School of Physics and Optoelectronic
Technology, Dalian University of Technology, Dalian 116023,
P.R.China}

\begin{abstract}
The generation of atomic entanglement is discussed in a system
that atoms are trapped in separate cavities which are connected
via optical fibers. Two distant atoms can be projected to
Bell-state by synchronized turning off the local laser fields and
then performing a single quantum measurement by a distant
controller. The distinct advantage of this scheme is that it works
in a regime that $\Delta\approx\kappa\gg g$, which makes the
scheme insensitive to cavity strong leakage. Moreover, the
fidelity is not affected by atomic spontaneous emission.
\end{abstract}
\maketitle

Very recently, much attention has been paid to the study of the
possibility of quantum information processing realized via optical
fibers $^{[1,2]}$. Generating an entangled state of distant qubits
turns out to be a basic aim of quantum computation. It has been
pointed out that implementing quantum entangling gate that works
for spatially separated local processors which are connected by
quantum channels is crucial in distributed quantum computation.
Many schemes have been put forward to prepare engineering
entanglement of atoms trapped in separate optical cavities by
creating direct or indirect interaction between them $^{[3-10]}$.
Some of the schemes involve direct connection of separate cavities
via optical fibers, others apply detection of the photons leaking
from the cavities. All the implemented quantum gates work in a
probabilistic way. To improve the corresponding success
probability and fidelity, one must construct precisely controlled
coherent evolutions of the global system and weaken the affect of
photon detection inefficiency. In the system considered by
Serafini et al $^{[5]}$, the only required local control is
synchronized switching on and off of the atom-field interaction in
the distant cavities. In the scheme proposed by Mancini and Bose
$^{[11]}$, a direct interaction between two atoms trapped in
distant cavities is engineered, the only required control for
implementing quantum entangling gate is turning off the
interaction between atoms and the locally applied laser fields. In
the present letter, we propose an alternative scheme with
particular focus on the establishment of three-qubit entanglement,
which is suitable and effective for the generation of three-atom
W-type state and two-atom Bell-state. To generate three-atom
W-type state, the only control required is synchronized turning
off the locally applied laser fields. While, To generate two-atom
Bell-state, an additional quantum measurement performed on one of
the atoms is needed. We demonstrate that the scheme works in a
high success probability, and the atomic spontaneous emission does
not affect the fidelity.

The schematic setup of the system is shown in Fig. 1. Three
two-level atoms 1, 2 and 3 locate in separate optical cavities
$C_{1}$, $C_{2}$ and $C_{3}$ respectively. The cavities are
assumed to be single-sided. Three off-resonant driving external
fields $\varepsilon _{1}$, $\varepsilon _{2}$ and $\varepsilon
_{3}$ are added on $C_{1}$, $C_{2}$ and $C_{3}$ respectively. In
each cavity, a local weak laser field is applied to resonantly
interact to the atom. Two neighboring cavities are connected via
optical fiber. The global system is located in vacuum.
\begin{figure}
\epsfig{file=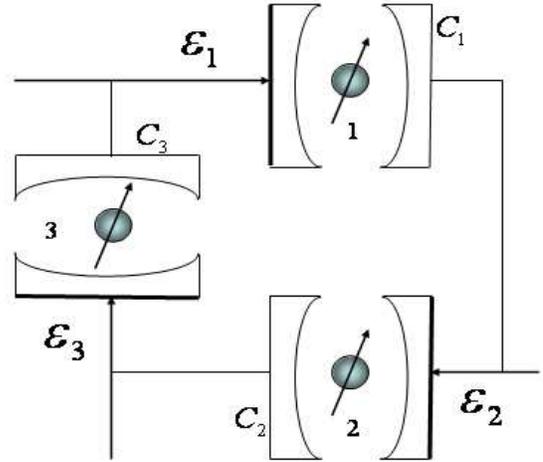, width=7.5cm,
height=6.5cm,bbllx=30,bblly=0,bburx=414,bbury=354}\caption{{\protect\footnotesize
{Schematic setup of the supposed system. Three two-level atoms are
trapped in separate optical cavities, which are connected via
optical fibers in turn. All the cavities are assumed to be
single-sided. Each of the cavities is driven by an external field.
Every atom is coupled to a local laser field.}}}
\end{figure}
Using the input-output theory, taking the adiabatic approximation
$^{[12]}$ and applying the methods developed in Refs. [11] and
[13], we obtain the effective Hamiltonian of the global system as
\begin{eqnarray}
H_{eff}=J_{12}\sigma _{1}^{z}\sigma _{2}^{z}+J_{23}\sigma
_{2}^{z}\sigma _{3}^{z}+J_{31}\sigma _{3}^{z}\sigma
_{1}^{z}+\Gamma\sum\limits_{i}(\sigma _{i}^{-}+\sigma _{i}^{+}),
\end{eqnarray}
where $\sigma_{i}^{z}$ and $\sigma_{i}^{+} (\sigma_{i}^{-})$,
$i=1,2,3$, are spin and spin raising (lowering) operators of atom
$i$, $\Gamma$ represents the local laser field added on the atom.
To keep the validity of adiabatic approximation, we assume $\Gamma
\ll J_{12}(J_{23},J_{31})$. And
\begin{eqnarray}
J_{12} &=&2\kappa\chi ^{2}Im\left\{ \alpha _{1} \alpha _{2} ^{\ast
}(Me^{i\phi
_{21}}+\kappa e^{i\phi_{32}+\phi_{13}})/(M^{3}-W^{3})\right\},  \nonumber \\
J_{23} &=&2\kappa \chi ^{2}Im\left\{ \alpha _{2} \alpha _{3}
^{\ast }(Me^{i\phi
_{32}}+\kappa e^{i\phi_{13}+\phi_{21}})/(M^{3}-W^{3})\right\},  \nonumber \\
J_{31} &=&2\kappa \chi ^{2}Im\left\{ \alpha _{3} \alpha _{1}
^{\ast }(Me^{i\phi _{13}}+\kappa
e^{i\phi_{21}+\phi_{32}})/(M^{3}-W^{3})\right\},
\end{eqnarray}
where $\kappa$ is the cavity leaking rate,
$\chi=\frac{g^2}{\Delta}$, $g$ is the coupling strength between
atom and cavity field, $\Delta$ is the detuning. In deducing Eq.
(1), the condition $\Delta\approx\kappa\gg g$ is assumed,
$M=i\Delta+\kappa$, $W^{3}=\kappa ^{3}e^{i(\phi _{21}+\phi
_{32}+\phi _{13})}$. The phase factors $\phi _{21}$, $\phi _{32}$,
and $\phi _{13}$ are the phases delay caused by the photon
transmission along the optical fibers. And
\begin{eqnarray}
\alpha_{1}&=&\frac{M^2\varepsilon_{1}+\kappa^{2}e^{i(\phi_{32}+\phi_{13})}
\varepsilon_{2}+M\kappa e^{i\phi_{13}}\varepsilon_{3}}{M^{3}-W^3},
\nonumber\\
\alpha_{2}&=&\frac{M^2\varepsilon_{2}+\kappa^{2}e^{i(\phi_{13}+\phi_{21})}
\varepsilon_{3}+M\kappa e^{i\phi_{21}}\varepsilon_{1}}{M^{3}-W^3},
\nonumber\\
\alpha_{3}&=&\frac{M^2\varepsilon_{3}+\kappa^{2}e^{i(\phi_{21}+\phi_{32})}
\varepsilon_{1}+M\kappa e^{i\phi_{32}}\varepsilon_{2}}{M^{3}-W^3},
\end{eqnarray}
 We assume that
$\varepsilon_{1}=\varepsilon_{2}=\varepsilon_{3}=\varepsilon_{0}$,
$\phi _{21}=\phi _{32}=\phi _{13}=\phi _{0}$. This leads to
\begin{eqnarray}
\alpha_{1}=\alpha_{2}=\alpha_{3}=\alpha_{0},\nonumber\\
J_{12}=J_{23}=J_{31}=J_{0}.
\end{eqnarray}
The Hamiltonian in Eq. (1) is now written as
\begin{eqnarray}
H_{eff}=H_{zz}+H_{x},
\end{eqnarray}
where
\begin{eqnarray}
H_{zz}=J_{0}(\sigma _{1}^{z}\sigma _{2}^{z}+\sigma _{2}^{z}\sigma
_{3}^{z}+\sigma _{3}^{z}\sigma _{1}^{z}),
H_{x}=\Gamma\sum\limits_{i}(\sigma _{i}^{-}+\sigma _{i}^{+}).
\end{eqnarray}
Eq. (5) represents the Hamiltonian of an Ising ring model. The
entanglement of the ground state of the above Hamiltonian has
already been discussed $^{[14]}$. Here, we study the entanglement
of the evolved system state governed by the Hamiltonian. Under the
condition $\Gamma\ll J_{0}$, the secular part of the effective
Hamiltonian can be obtained through the transformation
$UH_{x}U^{-1}$, $U=e^{-iH_{zz}t}$, as $^{[15]}$
\begin{eqnarray}
\tilde{H}=\frac{\Gamma}{2}[\sigma_{1}^{x}(1-\sigma_{2}^{z}\sigma_{3}^{z})
+\sigma_{2}^{x}(1-\sigma_{3}^{z}\sigma_{1}^{z})
+\sigma_{3}^{x}(1-\sigma_{1}^{z}\sigma_{2}^{z})].
\end{eqnarray}
The straight forward interpretation of this Hamiltonian is: the
spin of an atom in the Ising ring flips \emph{if and only if} its
two neighbors have opposite spins.

For the initial states that one or two of the atoms are excited,
the system state is restricted within the subspace spanned by the
following basis vectors
\begin{eqnarray}
|\phi_{1}\rangle&=&|egg\rangle, |\phi_{2}\rangle=|eeg\rangle,
|\phi_{3}\rangle=|geg\rangle,\nonumber \\
|\phi_{4}\rangle&=&|gee\rangle, |\phi_{5}\rangle=|gge\rangle,
|\phi_{6}\rangle=|ege\rangle.
\end{eqnarray}
The Hamiltonian in Eq. (7) can be written as
\begin{eqnarray}
\tilde{H}=\left(\begin{array}{cccccc}
0 & \Gamma & 0 & 0 & 0 & \Gamma\\
\Gamma & 0 & \Gamma & 0 & 0 & 0\\
0 & \Gamma & 0 & \Gamma & 0 & 0\\
0 & 0 & \Gamma & 0 & \Gamma & 0\\
0 & 0 & 0 & \Gamma & 0 & \Gamma\\
\Gamma & 0 & 0 & 0 & \Gamma & 0
\end{array}\right).
\end{eqnarray}
The eigenvalues of the Hamiltonian can be obtained as
$E_{12}=\pm\Gamma, E_{34}=\pm\Gamma, E_{56}=\pm2\Gamma$, and the
corresponding eigenvectors are
\begin{eqnarray}
|\psi_{12}\rangle&=&\frac{1}{2}(-|\phi_{1}\rangle\mp|\phi_{2}\rangle\pm|\phi_{4}\rangle+|\phi_{5}\rangle),\nonumber
\\
|\psi_{34}\rangle&=&\frac{1}{2}(\pm|\phi_{1}\rangle\mp|\phi_{3}\rangle-|\phi_{4}\rangle+|\phi_{6}\rangle),\nonumber
\\
|\psi_{56}\rangle&=&\frac{1}{\sqrt{6}}(|\phi_{1}\rangle\pm|\phi_{2}\rangle+|\phi_{3}\rangle\pm|\phi_{4}\rangle+|\phi_{5}\rangle\pm|\phi_{6}\rangle).
\end{eqnarray}

For initial system state
$|\Psi(0)\rangle=\sum\limits_{i}c_{i}(0)|\phi_{i}\rangle$, the
evolving system state can be written as
$|\Psi(t)\rangle=\sum\limits_{i}c_{i}(t)|\phi_{i}\rangle$, where
the coefficients $c_{i}(t)$ are given by $^{[8]}$
\begin{eqnarray}
c_{i}(t)=\sum\limits_{j}[S^{-1}]_{ij}[Sc(0)]_{j}e^{-iE_{j}t},
\end{eqnarray}
where
$c(0)=[c_{1}(0),c_{2}(0),c_{3}(0),c_{4}(0),c_{5}(0),c_{6}(0)]^{T}$,
and $S$ is the $6\times6$ unitary transformation matrix between
eigenvectors and basis vectors.

Now we discuss the evolving system state for initial state that
only one atom is excited, that is,
$|\Psi(0)\rangle=|\phi_{1}\rangle$, which leads to
$c(0)=[1,0,0,0,0,0]^{T}$, we can obtain
\begin{eqnarray}
c_{1}(t)&=&\frac{2}{3}\textrm{cos}\Gamma
t+\frac{1}{3}\textrm{cos}2\Gamma
t,\nonumber\\
c_{2}(t)&=&-\frac{1}{3}\textrm{sin}\Gamma
t-\frac{1}{3}\textrm{sin}2\Gamma
t,\nonumber\\
c_{3}(t)&=&-\frac{1}{3}\textrm{cos}\Gamma
t+\frac{1}{3}\textrm{cos}2\Gamma
t,\nonumber\\
c_{4}(t)&=&\frac{2}{3}\textrm{sin}\Gamma
t-\frac{1}{3}\textrm{sin}2\Gamma
t,\nonumber\\
c_{5}(t)&=&-\frac{1}{3}\textrm{cos}\Gamma
t+\frac{1}{3}\textrm{cos}2\Gamma
t,\nonumber\\
c_{6}(t)&=&-\frac{1}{3}\textrm{sin}\Gamma
t-\frac{1}{3}\textrm{sin}2\Gamma t.
\end{eqnarray}
It should be noted that the Hamiltonian in Eq. (6) or Eq. (7)
remains invariant under the permutation of atoms 1, 2 and 3. We
also note that the initial state $|egg\rangle$ has exchange
symmetry for atoms 2 and 3. So, there is no doubt that
$c_{5}(t)\equiv c_{3}(t)$ and $c_{6}(t)\equiv c_{2}(t)$.

Eqs. (12) lead to an resolvable analyzing of the three-atom or
two-atom entanglement nature of the involving system state. The
entanglement of three-partite pure states can be measured by
intrinsic three-partite entanglement which is defined as $^{[16]}$
\begin{equation}
C_{abc}=C_{a(bc)}-C_{ab}^{2}-C_{ac}^{2},
\end{equation}
where $C_{a(bc)}$, which represents the tangle between a subsystem
$a$ and the rest of the global system (denoted as $b, c$), is
represented as
\begin{equation}
C_{a(bc)}=4\textrm{Det}\rho_{a}=2(1-\textrm{Tr}\rho _{a}^{2}).
\end{equation}
and $C_{ab}$ ($C_{ac}$) is the well known Concurrence that is used
for entanglement measurement of qubits $a$ and $b$ ($a$ and $c$)
$^{[17]}$
\begin{eqnarray}
C_{ab}=C(\rho_{ab})=\max \{0,\lambda _{1}-\lambda _{2}-\lambda
_{3}-\lambda _{4}\},
\end{eqnarray}
where $\rho_{ab}$ is the reduced density matrix of qubits $a$ and
$b$, $\lambda _{1}, \lambda_{2}, \lambda_{3}$ and $\lambda_{4}$
are four non-negative square roots of the eigenvalues of the
non-Hermitian matrix $\rho_{ab}(\sigma _{y}\otimes \sigma
_{y})\rho_{ab}^{\ast }(\sigma _{y}\otimes \sigma _{y})$ in
decreasing order.

The entanglement is described in Fig. 2. The solid line represents
three-atom intrinsic entanglement, the dotted line represents the
entanglement of atoms $2$ and $3$, and the dashed line represents
the tangle between atom 2 and the rest two atoms.

\begin{figure}
\epsfig{file=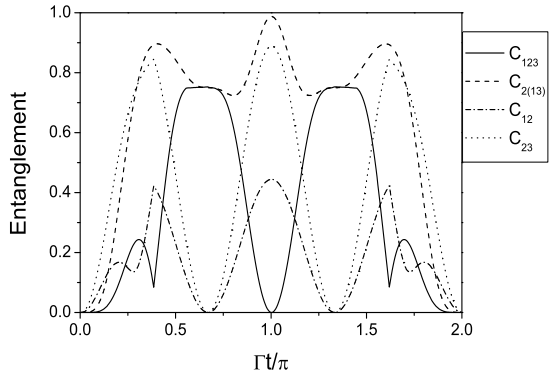, width=8cm,
height=6cm}\caption{{\protect\footnotesize {Entanglement of atoms
as a function of time (in units $\pi\Gamma^{-1}$) for three-atom
entanglement (solid line), tangle of atom 2 and the rest two atoms
(dashed line), entanglement of atom 1 and atom 2 (dash dotted
line), and entanglement of atom 2 and atom 3 (dotted line).}}}
\end{figure}

In most region of the time interval $(0,2)$, the tangle between
atom 2 and the rest two atoms does not alter too much. It seems
that two-atom entanglement makes the largest contribution to the
variety of three-atom entanglement, since the three-atom
entanglement is expressed by the difference between the tangle and
the two-atom entanglement [see Eq. (13)]. The peak entanglement of
atoms 2 and 3 is much larger than that of atoms 1 and 2. These may
suggest the following physical picture: for the initial state only
one atom is excited, the interaction between distant atoms can
generate strong and relatively steady entanglement shared by one
atom and the rest. Also, the interaction can cause strong
entanglement shared by any two atoms, while only the atoms that
are initially in ground state share the largest two-atom
entanglement.

In detail, two-atom entanglements $C_{12}$ and $C_{23}$ approach
their maximum at $\Gamma t_{1}=(2k+1)\pi$ ($k=0,1,2,3\cdots$),
where three-atom entanglement $C_{123}$ turns out to be zero. It
is clearly shown in Fig. 2 that $C_{23}$ is always larger than
$C_{12}$ in the whole region. In fact, it can be analytically
proved that $C_{23}=2C_{12}$ at $\Gamma t_{1}$. At $\Gamma
t_{2,3}=(2k+1)\pi\pm\frac{1}{3}\pi$, three-atom entanglement
$C_{123}$ periodically reaches a maximum, the corresponding
two-atom entanglement is zero.

At the points, the initial state evolves into the following states
\begin{eqnarray}
|\Psi(t_{1})\rangle&=&-\frac{1}{3}|egg\rangle+\frac{2\sqrt{2}}{3}|\Psi_{123}\rangle,\nonumber
\\
|\Psi(t_{2,3})\rangle&=&-\frac{1}{2}|egg\rangle\mp\frac{\sqrt{3}}{2}|gee\rangle,
\end{eqnarray}
where
$|\Psi_{123}\rangle=|g\rangle_{1}(|eg\rangle_{23}+|ge\rangle_{23})/\sqrt{2}=|g\rangle_{1}|\Psi^{+}\rangle_{23}$.
We can name $|\Psi_{123}\rangle$ as a Bell-correlated state.

In Eq. (16), $|\Psi(t_{1})\rangle$ is a combination of the initial
state and a Bell-correlated state, also it is a W-type state.
$|\Psi(t_{2,3})\rangle$ are linear combinations of the initial
state and a state with atomic population inverse with respect to
the initial state.

The results imply possible applications in practical distant
quantum communication. For example, it can be applied in the
preparation of maximally entangled state of distant atoms, and
thus acts as an atomic entangling gate. In this case, we assume
Alice, Bob, and Charles hold atoms 1, 2, and 3 respectively.

To do this, Alice, Bob, and Charles synchronously turn off their
locally applied laser fields at $t_{1}$. Now, they together have a
W-type state $|\Psi(t_{1})\rangle$. Then Alice performs
measurement $\sigma^{z}$ on her atom. She finds her atom is in
ground state with probability $\frac{8}{9}$, which is exactly the
success probability that the atoms held by Bob and Charles are
projected to Bell-state $|\Psi^{+}\rangle_{23}$, or she finds the
system state is recovered to initial state $|egg\rangle$ with
probability $\frac{1}{9}$.

The advantage of the scheme is that both Bob and Charles do not
need any measurement to entangle their atoms. All the requirement,
after the locally applied laser fields are turned off, is a
$\sigma^{z}$ measurement performed by Alice. So, two-atom
maximally entangled state can be generated by remote operation.
Especially, the measurement performed by Alice does not damage the
initial state of the global system if she failed to entangle the
others' atoms. Here, Alice can be regarded as a distant
controller, and her atom turns out to be a control-qubit.

In this process, the main obstacle is the spontaneous emission of
the atoms and the leakage of optical fibers.

We firstly investigate the affection of atomic spontaneous
emission. The evolution of the global system is now described by
the non-Hermitian conditional Hamiltonian $H_{s}=-i\gamma\sum
\limits_{i}|e\rangle_{i}\langle e|+\tilde{H}$ $^{[3]}$, where
$\gamma$ denotes the atomic spontaneous emission rate.

\begin{figure}
\epsfig{file=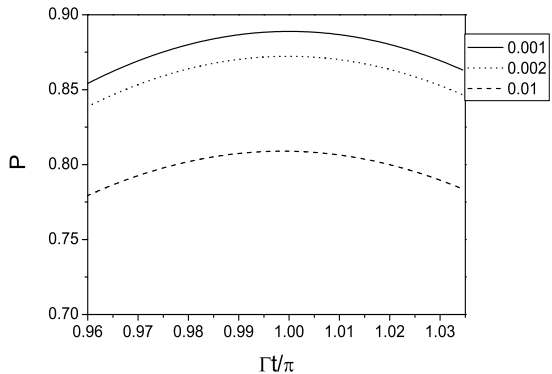, width=8cm,
height=6cm}\caption{{\protect\footnotesize {Success probability of
preparing Bell-state $|\Psi^{+}\rangle_{23}$ as a function of time
(in units $\pi\Gamma^{-1}$) for atomic spontaneous emission rates
$\gamma=0.001\Gamma$ (solid line), $\gamma=0.002\Gamma$ (dotted
line), and $\gamma=0.01\Gamma$ (dashed line).}}}
\end{figure}
In Fig. 3, we plot the success probability $P$ of preparing
Bell-state $|\Psi^{+}\rangle_{23}$ with respect to time for
different atomic spontaneous emission rates: $\gamma=0.001\Gamma$,
$\gamma=0.002\Gamma$, and $\gamma=0.01\Gamma$. The success
probability is undoubtedly sensitive to the atomic spontaneous
emission. The maximum probability drops to $0.881$, $0.872$, and
$0.809$ respectively. However, for any $\gamma$, the corresponding
fidelity can not be affected since $c_{3}(t)\equiv c_{5}(t)$
(recall that both the Hamiltonian and the initial state remain
invariant under the permutation of atom 2 and atom 3).

The dissipation of the photon leakage along optical fibers can be
included in the spin-spin coupling coefficients by the exchange
$e^{i\phi _{ij}}\rightarrow e^{i\phi _{ij}-\nu L}$, where $\nu$ is
the decay per meter and $L$ is the length of the optical fiber
between atoms $i$ and $j$. For typical fibers $^{[18]}$, the decay
per meter is $\nu=0.08$. The spin-spin coupling coefficient is now
about $90\%$ of that in Eq. (6). The adiabatic approximation
$\Gamma \ll J_{0}$ can still be fully kept. So the entangling gate
still works with high fidelity.

Another dissipation is the cavity leakage. While, in the adiabatic
approximation, we have assumed $\Delta \approx \kappa \gg g$. The
entanglement is then insensitive to the variety of strong leakage
rate.

In summary, we have discussed the remote generation of atomic
entanglement in a system contains three distant atoms for the
initial state that only one atom is excited. The atoms that are
initially in ground state share the largest two-atom entanglement.
Two-atom entanglement turns out to be the largest contribution to
the variety of three-atom entanglement. In an application of
preparing entangled state of two atoms, a quantum measurement of
$\sigma^{z}$ performed on the atom that is initially excited at
typical time is required after synchronized turning off the
locally applied laser fields. The success probability that two
atoms are prepared in Bell-state $|\Psi^{+}\rangle _{23}$ can
approach $\frac{8}{9}$. The distinct advantage of this scheme lies
in the large detuning and large cavity leakage, that is
$\Delta\approx\kappa\gg g$ which loosens the requirement of cavity
dissipation. Furthermore, we show that the fidelity of the scheme
is not affected by the atomic spontaneous emission. We think this
scheme may work as a candidate for scalable long-distance quantum
communication or one-way quantum computation $^{[3]}$.

This work is supported by NSF of China under Grant Nos. 10647107
and 10575017.

\end{document}